\documentclass[aip,jcp,preprint,eqsecnum,a4paper]{revtex4-1}

\usepackage{graphicx}
\usepackage[sort&compress]{natbib}
\usepackage{braket}
\usepackage{amsfonts}
\usepackage{amssymb}
\usepackage{dcolumn}
\usepackage{subdepth}
\usepackage{color}
\usepackage{enumitem}
\usepackage{blindtext}
\usepackage{soul}
\usepackage{hyperref}
\usepackage{acro}

\DeclareAcronym{DOF}{
 short = DOF ,
 long  = degree of freedom,
 long-plural-form = degrees of freedom,
 class = abbrev
}

\DeclareAcronym{GLE}{
 short = GLE ,
 long  = Generalized Langevin Equation,
 class = abbrev
}

\DeclareAcronym{CL}{
 short = CL ,
 long  = Caldeira-Leggett,
 class = abbrev
}

\DeclareAcronym{EOM}{
 short = EOM ,
 long  = equation of motion,
 long-plural-form = equations of motion,
 class = abbrev
}

\DeclareAcronym{FDT}{
 short = FDT ,
 long  = fluctuation-dissipation theorem,
 class = abbrev
}

\DeclareAcronym{MD}{
 short = MD,
 long  = molecular dynamics,
 class = abbrev
}

\DeclareAcronym{HEOM}{
 short = HEOM,
 long  = hierarchy equations of motion,
 class = abbrev
}

\DeclareAcronym{HOPS}{
 short = HOPS,
 long  = hierarchy of pure states,
 class = abbrev
}

\DeclareAcronym{SLN}{
 short = SLN,
 long  = stochastic Liouville-von Neumann equations,
 class = abbrev
}

\DeclareAcronym{PI}{
 short = PI ,
 long  = path integral,
 class = abbrev
}

\DeclareAcronym{MAF}{
 short = MAF,
 long  = momentum autocorrelation function,
 class = abbrev
}

\DeclareAcronym{MFC}{
 short = MFC,
 long  = momentum-force correlation function,
 class = abbrev
}

\DeclareAcronym{MCTDH}{
 short = MCTDH,
 long  = multi-configurational time-dependent Hartree,
 class = abbrev
}

\DeclareAcronym{HK}{
 short = HK,
 long  = Herman-Kluk,
 class = abbrev
}

\DeclareAcronym{IVR}{
 short = IVR,
 long  = initial-value representation,
 class = abbrev
}

\DeclareAcronym{LSC}{
 short = LSC-IVR,
 long  = linearized semiclassical initial-value representation,
 class = abbrev
}

\DeclareAcronym{TGWD}{
 short = TGWD,
 long  = thawed Gaussian wave packet dynamics,
 class = abbrev
}

\newcommand{\DOF}{\ac{DOF}}
\newcommand{\DOFs}{\acp{DOF}}
\newcommand{\GLE}{\ac{GLE}}

\newcommand{\CL}{\ac{CL}}
\newcommand{\EOM}{\ac{EOM}}
\newcommand{\EOMs}{\acp{EOM}}
\newcommand{\HEOM}{\ac{HEOM}}
\newcommand{\HOPS}{\ac{HOPS}}
\newcommand{\SLN}{\ac{SLN}}

\newcommand{\FDT}{\ac{FDT}}
\newcommand{\MD}{\ac{MD}}
\newcommand{\MAF}{\ac{MAF}}
\newcommand{\MFC}{\ac{MFC}}
\newcommand{\PI}{\ac{PI}}

\newcommand{\HK}{\ac{HK}}
\newcommand{\IVR}{\ac{IVR}}

\newcommand{\LSC}{\ac{LSC}}
\newcommand{\TGWD}{\ac{TGWD}}
\newcommand{\MCTDH}{\ac{MCTDH}}

\newcommand{\VS}{V_{\mathrm{S}}}

\newcommand{\Eq}[1]{Eq.~(\ref{#1})}

\newcommand{\Fig}[1]{Fig.~\ref{#1}}
\newcommand{\Sec}[1]{Sec.~\ref{#1}}

\newcommand{\diff}{\mathrm{d}}
\renewcommand{\i}{\mathrm{i}}
\newcommand{\e}{\mathrm{e}}

\newcommand{\num}{^{(\mathrm{num})}}

\bibpunct{[}{]}{,}{n}{}{;}

\begin{document}

\title{On Computing Spectral Densities from Classical, Semiclassical and Quantum Simulations}

\author{Fabian Gottwald}
\author{Sergei D. Ivanov}
\email{sergei.ivanov@uni-rostock.de}
\author{Oliver K\"uhn}
\affiliation{Institute of Physics, University of Rostock,
  Albert Einstein Stra{\ss}e 23-24, 18059 Rostock, Germany}

\begin{abstract}
The Caldeira-Leggett model provides a compact characterization of a thermal environment in terms of a spectral density function.
This simplicity has led to a variety of numerically exact quantum methods for reduced density matrix propagation.
When using these methods, a spectral density has to be computed from dynamical properties of system and environment, which
is commonly done using classical molecular dynamics simulations.
However, there are situations, where quantum effects could play a role.
Therefore, we reformulate our recently developed Fourier method in order to enable spectral density calculations from semiclassical simulations which approximately consider quantum effects.
We propose two possible protocols based on either correlation functions or expectation values.
These protocols are tested for the linearized semiclassical initial-value representation (LSC-IVR), the thawed Gaussian wave packet dynamics (TGWD) and hybrid schemes combining the two with the more accurate Herman-Kluk (HK) formula.
Surprisingly, spectral densities from the LSC-IVR method, based on a completely classical propagation, are extremely accurate whereas those from the single-trajectory TGWD are of poor quality in the anharmonic regime.
The hybrid methods provide reasonable quality when the system is close to the classical regime, although, at finite temperature, the computation protocol from expectation values turns out to be more robust.
If stronger quantum effects are observed, both hybrid methods turn out as too inaccurate.

\end{abstract}

\date{\today}

\maketitle

\section{Introduction}
One of the main research goals in chemical physics is to obtain a comprehensive understanding of processes in molecular systems, that are usually influenced by their environments~\cite{Kuehn2011}.
Unraveling these processes and deducing the underlying basic mechanisms and timescales requires the interplay of sophisticated experimental techniques and reliable theoretical models.
Modern computer facilities allow theoretical physicists to simulate an increasingly large number of interacting \DOFs\ using a variety of methods.
%
In many cases, however, a reduction of the complexity to a few relevant \DOFs, termed system, is not only helpful for the interpretation but also opens a doorway to a whole palette of approximations for treating the irrelevant \DOFs, termed bath, on a lower accuracy level~\cite{Weiss-Book-2012}.
In the context of such a system-bath partitioning, a convenient approach is to map the usually high-dimensional bath onto a simple model.
This allows one to derive so-called reduced \EOMs\ for the system \DOFs\ in which the influence of the bath is accounted for implicitly~\cite{Kuehn2011,Weiss-Book-2012, ZwanzigBook2001}.
A very popular system-bath model is the \CL\ one, in which the bath is mimicked by a collection of harmonic oscillators bi-linearly coupled to the system coordinates~\cite{Zwanzig1973,Caldeira-PRL-1981,Caldeira-AP-1983}.
In the corresponding reduced \EOM, the so-called \GLE, the influence of the bath is limited to dissipation and fluctuations only, whose properties can be described by a single spectral density function.
This compact characterization of the bath enables a full quantum-dynamical treatment of the reduced density matrix, since the remaining system part is usually of low dimensionality.
In particular, real-time path integral techniques, such as \HEOM~\cite{Tanimura-JPhysSocJpn-1989,Tanimura-JPhysSocJpn-2005,Tanimura2006}, can be derived from the Feynman-Vernon influence functional approach~\cite{Feynman-AP-1963} and have become routine in many applications in condensed phase spectroscopy~\cite{Mukamel1995, Tanimura2009} or exciton dynamics~\cite{Schroeter-PR-2015} to mention but two.
Further, stochastic approaches like \HOPS\ or \SLN\ have appeared and provided a significant gain in performance~\cite{Diosi-PRA-1998, Suess-PRL-2014, Stockburger2002, Stockburger2004}.
Finally, the \MCTDH\ method~\cite{meyer-MCTDH-2009} can be efficiently applied for an explicit propagation of system and bath oscillators due to the factorized potential form imposed by the \CL\ model.

In order to profit from the aforementioned quantum propagation methods, one needs to obtain a spectral density that reflects dynamical properties of the environment under study.
Typically, this is achieved via experimental spectroscopic measurements or by explicit simulations.
In addition, a variety of model spectral densities exists, however using model baths may yield qualitative understanding but cannot guarantee quantitative predictions.
When it comes to simulation, the dynamics should be ideally based on the full time-dependent Schr\"odinger equation which is, in practice, impossible for an arbitrary complex system due to the infamous 'exponential wall' stating an exponential increase of the work load with the number of \DOFs.
However, the spectral density can be universally defined from both classical and quantum-mechanical versions of the \GLE, which can be rigorously derived~\cite{ZwanzigBook2001,Weiss-Book-2012}.
This observation should allow one to use both fully quantum and purely classical \MD\ simulations for computing it, and, thereby, to establish the mapping of the system in question onto the \CL\ model.
It is worth pointing out that such a mapping can be either direct, that is the bath has the required form from the outset, or effective, meaning that the resulting spectral density properly mimics the influence of the bath on system properties at a price of losing the atomistic picture.
In case of a successful mapping, the spectral density should be well usable in the reduced quantum schemes which then enable a quantum treatment without any further approximations.
In this respect, the \CL\ model can be viewed as a promising framework for open dynamics connecting explicit classical \MD\ simulations with a corresponding reduced quantum-dynamical treatment based on spectral densities.
However, doubts about this mindset can arise from previous studies, where the \CL\ model has been shown to suffer from the so-called invertibility problem~\cite{Gottwald-JPCL-2015,Gottwald-JCP-2016}.
It states that a mapping of a realistic system onto a model is not invertible, thereby undermining the self-consistency and, thus, validity of the \CL\ model.
%
%
Since the independence of the spectral density on quantum effects only follows from a strictly valid \CL\ model, one can, for the general case, question the possibility to compute it from purely classical \MD\ simulations.
%
%
In case of the aforementioned effective mapping it might be even necessary to account for quantum effects in the dynamics as the system-bath coupling can be particularly sensitive to them, for instance, if quantum (de-)coherence, tunnelling, zero-point energy fluctuations or the discrete energy level structure significantly shape the processes.
In these cases, a full quantum-dynamical treatment of system and bath can not be avoided and reliable approximations in order to facilitate their propagation have to be found.

Promising candidates for approximations can be built on semiclassical techniques, which 
provide approximations to the quantum propagator, based on classical trajectories~\cite{VanVleck-PNAS-1928, Littlejohn1992}.
Quantum effects are accounted for by a phase factor, containing the classical action, as well as a prefactor constructed from classical monodromy matrix elements.
Since its initial formulation, the semiclassical theory has developed into a practical \IVR\ form, whose conceptually most convenient version is known as the \HK\ propagator~\cite{Heller-JCP-1981, Herman-Kluk-CP-1984, Herman-Kluk-JCP-1986}, see also reviews in Refs.~\cite{Miller-ACP-1974,Miller-JPCA-2001,Thoss-AR-2004,Kay-AR-2005}.
Although formulated in terms of classical objects, the \HK\ propagator is still of limited use due to the so-called sign problem caused by the rapid oscillations in the phase factor and the divergence of classical monodromy matrix elements with increasing dimensionality and non-linearity.
This makes the \IVR\ integral almost impossible to converge for larger but even moderate-sized molecular systems.
In order to improve the performance, several technical manipulations have been performed ranging from Filinov filtering~\cite{Makri-CPL-1987, Walton-MolPhys-1996,Herman-CPL-1997}, time-averaging methods~\cite{Elran-JCP-1999a, Elran-JCP-1999b,Kaledin-JCP-2003, Ceotto-JCP-2009} and improved sampling techniques~\cite{Tao-JCP-2009,Tao-JCP-2014} to the famous forward-backward schemes~\cite{Makri-CPL-1987, Walton-MolPhys-1996,Herman-CPL-1997} that have been used already in \PI\ techniques.
Other approaches apply more severe approximations to the \HK\ formula such as the \LSC, in which observables are treated in a purely classical fashion, but the dynamics is started from the quantum-mechanically correct initial state~\cite{Heller-JCP-1975, Miller-JCP-1998}.
Another approximation is \TGWD, which employs a Gaussian wave packet whose center and width are given in terms of a single trajectory and its monodromy matrix, respectively~\cite{Heller-JCP-1975}.
%
Although these methods are applicable to quite complex systems~\cite{Liu2009}, problems like zero-point energy leakage usually plague the \LSC\ propagation~\cite{Habershon-JCP-2009} and the \TGWD\ method is often lacking a correct description of quantum interference.
This gave rise to semiclassical hybrid schemes, which build a compromise between accuracy and performance and which are fully compatible with the system-bath partitioning idea.
While for the relevant system \DOFs\ the full \HK\ formula is maintained, the irrelevant bath \DOFs\ are treated by one of the aforementioned lower level approximations.
A hybrid scheme combining the \HK\ formula with the \TGWD\ method has been developed by Gro\ss mann and co-workers~\cite{Grossmann-JCP-2006, Goletz2009} and has been applied successfully in condensed phase dynamics and spectroscopy~\cite{Buchholz2012, Buchholz-JCP-2016}.
Further, combinations of the \HK\ formula with the \LSC\ method have been constructed~\cite{Sun-JCP-1997}, with the most recent contribution formulated in the Wigner representation by Koda~\cite{Koda-JCP-2016}.
Since these hybrid schemes appear to be especially tailored for the system-bath problem under study, we consider them as a suitable tool for computing spectral densities taking quantum effects approximately into account.

The main goal of this paper is to clarify how a spectral density should be calculated for a reduced quantum propagation.
In particular, it is checked whether a fully classical treatment is sufficient or whether exact or approximate quantum dynamics should be exploited. 
For the latter, it is investigated if the aforementioned semiclassical methods can provide an accurate way for computing spectral densities in the presence of dynamical quantum effects.

We start with a brief review of the theoretical basics of the \CL\ model in \Sec{sec: Theory and method}. 
Then, we reformulate our Fourier method, originally developed for computing spectral densities from classical equilibrium correlation functions~\cite{Gottwald-JCP-2015,Gottwald-JCP-2016}, according to the regime of quantum dissipation and outline two ways of using it in combination with quantum dynamical simulations. 
Then, the semiclassical techniques are reviewed in more detail.
In \Sec{sec: numerical tests} we present numerical results for different examples of quantum dynamics in a model heat bath.
Besides possible intrinsic semiclassical errors, the so-called 'invertibility problem' could be a conceptual problem rendering a spectral density useless, see Ref.~\cite{Gottwald-JPCL-2015}.
This problem states, that a mapping onto the \CL\ model is generally not invertible for anharmonic system potentials and, thus, dynamical quantities may be not correctly described.
Since this issue should be always investigated separately via a self-consistency check~\cite{Gottwald-JPCL-2015, Gottwald-JCP-2016}, we only analyze the semiclassical accuracy as such and impose a strict \CL\ model form of the bath according to a model spectral density.
This way, the invertibility problem is avoided from the outset and it can be tested whether the model spectral density is reproduced accurately if semiclassical propagation methods are used.
A summary with the main conclusions and an outlook to future studies is given in \Sec{sec: conclusions}.

\section{Theoretical Details and Methods}
\label{sec: Theory and method}

\subsection{The Caldeira-Leggett Model and the Spectral Density}
\label{sec: spectral density}
Within the \CL\ model, the system of interest comprises a selected (nuclear) \DOF\ $q$ with the mass $m$, which is bi-linearly coupled to a thermal heat bath represented as a collection of harmonic oscillators;~\cite{Caldeira-PRL-1981, Caldeira-AP-1983,Weiss-Book-2012,ZwanzigBook2001}
note that the system is chosen one-dimensional for the sake of presentation.
The full system and bath potential reads
\begin{eqnarray}
\label{eq: CL potential}
V(q,\{Q_i\}) & = &  \VS(q) + \sum_i \frac{1}{2}\omega_i^2 \left ( Q_i - \frac{g_i}{\omega_i^2} q \right )^2 \nonumber \\
& = & \VS(q) + \sum_i  \frac{1}{2}\omega_i^2 Q_i^2 - \sum_i g_i Q_i q + \sum_i \frac{1}{2}\frac{g_i^2}{\omega_i^2}q^2
\enspace ,
\end{eqnarray}
where $\{\omega_i\}$ and $\{Q_i\}$ denote the bath frequencies and mass-weighted coordinates, respectively.
The system potential $\VS(q)$ can be chosen arbitrarily and the system-bath coupling strength is regulated by the parameters $\{g_i\}$. 
The last term in the second line of \Eq{eq: CL potential} depends on the system coordinate only and constitutes a counter term in order to remove frequency renormalization effects and to make the full Hamiltonian translationally invariant, see the discussions in~\cite{Caldeira-AP-1983, Weiss-Book-2012}. 

Starting from the \CL\ model, one can derive the so-called \GLE~\cite{Zwanzig1973,Weiss-Book-2012,ZwanzigBook2001}, which is an \EOM\ for the system part
\begin{eqnarray}
\label{eq: GLE}
\dot q(t) & = & \frac{p(t)}{m} \nonumber \\
\dot p(t) & = & F[q(t)] - \intop_0^t \diff \tau \, \xi(t-\tau) p(\tau) + R(t) - m q(0) \xi(t)
\enspace ,
\end{eqnarray}
where $p$ is the momentum conjugate to $q$.
The total 
force in \Eq{eq: GLE} is
determined by the system force $F(q)=-\VS'(q)$, a non-Markovian friction term with the so-called memory kernel $\xi(t)$, a fluctuating force $R(t)$ and a
(rather artificial) term depending on the initial system coordinate~\cite{Weiss-Book-2012}.
Although often mimicked by a stochastic process, the fluctuating force can be explicitly expressed in terms of the free bath evolution $Q_i(t)$ as
\begin{eqnarray}
\label{eq: noise}
R(t) & = & \sum_i g_i Q_i(t)  \nonumber \\
 & = & \sum_i g_i \left [  \frac{P_i(0)}{\omega_i} \sin(\omega_i t) +  Q_i(0) \cos(\omega_i t) \right ] \enspace .
\end{eqnarray}
The \GLE\ can be understood either classically as an \EOM\ for the phase-space variables $(q,p)$ or quantum-mechanically as a Heisenberg \EOM\ for position and momentum operators~\cite{Weiss-Book-2012, Ford1985, Ford-PRA-1988}.
In both cases, the whole influence of the bath can be characterized by a single spectral density function $J(\omega)$ defined as the coupling-weighted distribution of bath frequencies
\begin{equation}
\label{eq: spectral density}
J(\omega)=\frac{\pi}{2} \sum_i \frac{g_i^2}{\omega_i} \delta(\omega - \omega_i) \enspace.
\end{equation}
In particular, the spectral density fully determines the memory kernel $\xi(t)$ via a cosine transform
\begin{eqnarray}
\label{eq: memory kernel}
\xi(t) & = & \frac{2}{m \pi} \intop_0^\infty \diff \omega \, \frac{J(\omega)}{\omega} \cos(\omega t) \nonumber \\
& = & \frac{1}{m} \sum_i \frac{g_i^2}{\omega_i^2} \cos(\omega_i t)
\end{eqnarray}
and the correlation function of the fluctuating force $R(t)$ via the so-called \FDT, whose classical and quantum-mechanical forms read
\begin{eqnarray}
\label{eq: FDT}
\langle R(0)R(t) \rangle_\mathrm{cl} & = & \frac{2 kT}{\pi} \intop_0^\infty \diff \omega \, \frac{J(\omega)}{\omega} \cos(\omega t) = mkT \xi(t) \nonumber \\
\langle R(0)R(t) \rangle_\mathrm{qm} & = & \frac{\hbar}{\pi} \intop_0^\infty \diff \omega \, J(\omega) \left [ \coth \left (  \frac{\hbar \omega}{2 kT} \right ) \cos(\omega t) - \i \sin(\omega t) \right ] \enspace .
\end{eqnarray}
Note that the fluctuations $R(t)$ become operator-valued in the quantum-mechanical case and are non-commutative at different times.
The averages $\langle ... \rangle$ exploited above are defined classically as a phase-space integral and quantum-mechanically as a trace over the initial system and bath states.
As it is typical for problems of quantum dissipative dynamics, this initial state is assumed to be a factorization of bath and system densities, i.e.\ $\rho = \rho_\mathrm S \cdot \rho_\mathrm B$, with an arbitrary (non-equilibrium) system density $\rho_\mathrm S$ and an equilibrium bath density 
\begin{equation}
\label{eq: bath density}
\rho_\mathrm B \equiv \frac{1}{Z_\mathrm B} \exp \left [- \frac{H_\mathrm B}{kT} \right ] 
\enspace ,
\end{equation}
with a finite temperature $T$, the partition function $Z_\mathrm B$ and the bath Hamiltonian 
\begin{equation}
H_\mathrm B = \sum_i \frac{P_i^2}{2} + \frac{1}{2}\omega_i^2 Q_i^2 \enspace .
\end{equation}
Note that this choice of the initial bath density directly implies a Gaussian statistics with $\langle R(t) \rangle = 0 $ for all times $t$.
The artificial term in the \GLE, \Eq{eq: GLE}, depending on the initial coordinate $q(0)$ is a direct consequence of this uncorrelated initial state and can be formally included into the noise term.
Importantly, if in \Eq{eq: bath density} the free bath Hamiltonian $H_\mathrm B$ is changed to
\begin{equation}
H'_\mathrm B = \sum_i \frac{P_i^2}{2} + \frac{1}{2}\omega_i^2 \left (Q_i - \frac{g_i^2}{\omega_i^2} q \right )^2 \enspace ,
\end{equation}
this newly defined noise obeys the same statistical properties as given by the \FDT\ in \Eq{eq: FDT}.
Nevertheless, we here follow closely the standard treatments of quantum dissipative systems,
which are typically based on the bath density in the form of \Eq{eq: bath density}.
Finally, it is worth stressing that the definition of the spectral density in \Eq{eq: spectral density} follows universally from the \GLE, irrespectively whether it is derived classically or quantum-mechanically.
%
The correct quantum statistics in the \FDT\, see second line of \Eq{eq: FDT}, is manifested by the hyperbolic cotangent, which is not a part of the spectral density but rather follows from using the quantum-statistical bath density operator, i.e.\ \Eq{eq: bath density}, as the initial state.

\subsection{The Fourier Method}
Recently, we have established a Fourier-domain protocol to parameterize spectral densities from explicit classical \MD\ simulations.~\cite{Gottwald-JPCL-2015,Gottwald-JCP-2016}
In order to develop a similar procedure based on the 
%
%
\GLE\ in \Eq{eq: GLE}, one can proceed in two different ways.
One way is to follow the same pathway as in Refs.~\citenum{Gottwald-JPCL-2015,Gottwald-JCP-2016} and to multiply the \GLE\ by the initial momentum $p(0)$ followed by an average $\langle ... \rangle$ with respect to the initially factorized system and bath states.
Irrespectively of whether a classical or quantum average is performed, this amounts to an integro-differential equation in terms of the \MAF\ $C_{pp}(t)\equiv\langle p(0)p(t) \rangle$ and the \MFC\ $C_{pF}(t)\equiv\langle p(0)F(t) \rangle$
\begin{eqnarray}
\label{eq: eq for Cpp}
\dot C_{pp}(t) & = & C_{pF}(t) - \intop_0^t \diff \tau \, \xi(t-\tau) C_{pp}(\tau) - m \langle p q\rangle \xi(t)
\enspace ,
\end{eqnarray}
where the absence of correlation between the noise $R(t)$ and initial momentum has been used.
%
The only difference to the working equation in Ref.~\onlinecite{Gottwald-JCP-2016} is the presence of the last term on the right hand side.
The average $ \langle p q \rangle \equiv \langle p(0) q(0) \rangle$ therein follows from the artificial term in \Eq{eq: GLE} and represents the initial correlation between position and momentum, which usually vanishes in the classical case.
In the quantum case, however, $p$ and $q$ are Hermitian operators with special commutation relations leading to non-vanishing and purely imaginary correlations.
The integro-differential equation in \Eq{eq: eq for Cpp} is now transformed into the frequency domain via a half-sided Fourier transform, i.e.\ $\hat \bullet = \intop_0^\infty \e^{-\i \omega t} \, \bullet$, with hats denoting Fourier-transformed functions.
The convolution thereby turns into a product $\hat \xi(\omega) \cdot \hat C_{pp}(\omega)$ and the time derivative $\dot C_{pp}(t)$ of the \MAF\ transforms into $\i \omega \hat C_{pp}(\omega) - C_{pp}(t=0)$ with $C_{pp}(t=0)= \langle p^2 \rangle $ being the second moment of the initial momentum.
Overall, one obtains an algebraic equation that can be solved for the Fourier-transformed
memory kernel $\hat \xi(\omega)$
\begin{equation}
\label{eq: xi from TCF}
\hat \xi(\omega) =  \frac{\hat C_{pF}(\omega) + \langle p^2 \rangle -\i \omega \hat C_{pp}(\omega)}{\hat C_{pp}(\omega) + m \langle pq \rangle} \enspace .
\end{equation}
The memory kernel in frequency domain is now expressed in terms of the Fourier-transformed time-correlation functions $\hat C_{pp}(\omega)$ and $\hat C_{pF}(\omega)$ and the connection to the spectral density is given by its real part
\begin{equation}
\label{eq: spectral density and xi}
J(\omega)=m \omega \, \mathrm{Re} \, \hat \xi(\omega) \enspace ,
\end{equation}
which can be verified using the basic definitions in \Sec{sec: spectral density}.

Another equation can be derived along the same lines but starting from averaging the \GLE\ itself without multiplying it with the initial momentum first.
This results in an integro-differential equation in terms of the momentum $\langle p(t) \rangle $ and system force $\langle F(t) \rangle $ expectation values
\begin{eqnarray}
\langle \dot p(t) \rangle & = & \langle F(t) \rangle - \intop_0^t \diff \tau \, \xi(t-\tau) \langle p(\tau) \rangle- m \langle q \rangle \, \xi(t) \enspace ,
\end{eqnarray}
where it has been used that the noise term $R(t)$ has a zero mean.
The last term now includes the average of the initial position $\langle q \rangle \equiv \langle q(0) \rangle$ taken with respect to the initial system state $\rho_\mathrm S$.
In frequency domain, the resulting expression for $\hat \xi(\omega)$ reads
\begin{equation}
\label{eq: xi from averages}
\hat \xi(\omega) = \frac{ \langle \hat F(\omega) \rangle + \langle p \rangle -\i \omega \langle \hat p(\omega) \rangle }{\langle \hat p(\omega) \rangle + m \langle q \rangle} \enspace ,
\end{equation}
with the average $\langle p \rangle$ of the initial momentum and the Fourier-transformed expectation values $ \langle \hat p(\omega) \rangle $ and $\langle \hat F(\omega) \rangle $.

The two formulas, i.e.\ \Eq{eq: xi from TCF} and \Eq{eq: xi from averages}, can be directly used to map a given system coupled to a complex environment onto the \CL\ model form, provided that such a mapping is self-consistent~\cite{Gottwald-JPCL-2015}.
For this purpose one needs to compute the dynamical quantities from an explicit propagation of the system in its environment and insert them into corresponding equations in order to obtain the memory kernel and, thereby, the spectral density via \Eq{eq: spectral density and xi}.
Importantly, the derivation presented above is valid also in the quantum regime and, hence, quantum effects in the dynamical quantities can be considered.
Finally, we would like to stress that typical quantum correction factors, usually occurring in quantum bath correlation functions for electronic transitions~\cite{Ritschel2014}, need \textit{not} be employed in the aforementioned equations.
In contrast, the correct statistical properties of the quantum fluctuations are implicitly encoded in the correlation functions or expectation values, which should be computed via a quantum-dynamical simulation started from a quantum-mechanically thermalized initial bath state.

\subsection{Semiclassical Simulation Techniques}
\label{sec: semiclassics}
In the following,
the semiclassical techniques are reviewed that provide a way to approximately incorporate quantum effects into the dynamical quantities needed for spectral density calculations.
Here, only the ideas behind the semiclassical propagators and their basic structure are presented, whereas all expressions for the dynamical quantities of interest are shifted to the supplement.
The starting point is the so-called \HK\ propagator developed by Heller and later extended by Herman and Kluk~\cite{Heller-JCP-1981, Herman-Kluk-CP-1984, Herman-Kluk-JCP-1986}.
This propagator can be formulated either in Hilbert space for the density matrix or in the Wigner representation for evolving Wigner functions~\cite{Koda-JCP-2015}.
However, in a recent study we have shown that the two formulations are fully equivalent both from the algebraic and the numerical perspective~\cite{Gottwald-CP-2018}.
Since the common approximations discussed below are most clearly expressed in the Wigner representation, it is employed in the following.

For a general $f$-dimensional system, the 
time evolution of a Wigner function is provided by the
\HK\ propagator applied to an initial Wigner function $W_0(z)$~\cite{Koda-JCP-2015}
\begin{eqnarray}
\label{eq: Wigner HK}
W_{\mathrm{HK}}^{}(z,t) &  = &   \displaystyle \intop \frac{\diff \bar z_0 \, \diff \Delta z_0}{(2 \pi \hbar)^{2f} } \, \tilde C_t(\bar z_0,\Delta z_0) \, \e^{\frac{\i}{\hbar} \tilde S_t(\bar z_0,\Delta z_0)} g(z ;\bar z_t, \Delta z_t) \nonumber \\
& & \times \intop \diff z' g^*(z'; \bar z_0, \Delta z_0) W_0( z') \enspace ,
\end{eqnarray}
where the notation $z\equiv(q,p)$ is used for a point in the classical phase space. 
This expression contains an integral over initial midpoints $\bar z_0 \equiv \frac{1}{2} (z_0^+ + z_0^-)$ and differences $\Delta z_0 \equiv (z_0^+ - z_0^-)$ of a pair of classical trajectories $z_t^\pm$, the action 
\begin{equation}
\label{eq: action}
\tilde S_t(\bar z_0, \Delta z_0)=\intop_0^t \diff \tau  \left [ \dot {\bar z}_\tau ^T \mathbf J \Delta z_\tau - H^+(z^+_\tau) + H^-(z^-_\tau) \right ] \enspace ,
\end{equation}
with the symplectic matrix 
\begin{equation}
 \mathbf J= \left (
 \begin{array}{cc}
   0 & 1\\
  -1 & 0\\
 \end{array} \right )
 \enspace ,
\end{equation}
and the classical Hamilton functions $H^\pm(z^\pm)$, which generate the classical propagation according to Hamilton equations.
The functions $g(z;\bar z, \Delta z)$ can be interpreted as a generalization of coherent states to the  phase space and constitute Gaussian functions located at the midpoints of the trajectory pair 
\begin{eqnarray}
\label{eq: coherent state phase space}
g(z;\bar z, \Delta z) = \det \left ( \frac{\Gamma}{\pi \hbar } \right )^{1/4} \exp\left[ - \frac{1}{2 \hbar} (z-\bar z)^T \Gamma (z- \bar z) + \frac{\i}{\hbar}  \Delta z^T \mathbf J^T (z-\bar z)\right] 
\enspace,
\end{eqnarray}
with a positive definite $(2f \times 2f)$-matrix $\Gamma$.
Finally, a prefactor $\tilde C_t(\bar z_0, \Delta z_0)$ given in terms of the classical monodromy matrices $\mathbf M_t^\pm$ reads
\begin{eqnarray}
\label{eq: prefactor phase space}
\tilde C_t(\bar z_0, \Delta z_0) & = &(\det \Gamma)^{-1/2}   \det \left[ \frac{1}{2}\left ( \frac{ \Gamma}{2} + \i \mathbf J \right ) \mathbf M_t^+ \left ( 1+  \i \mathbf J  \frac{ \Gamma}{2} \right) \right.\nonumber \\ 
&+&  \left. \frac{1}{2} \left ( \frac{ \Gamma}{2} - \i \mathbf J \right) \mathbf M^-_t \left ( 1-  \i \mathbf J  \frac{ \Gamma}{2}   \right)  \right]^{1/2}
\enspace .
\end{eqnarray}
The \HK\ formula has proven itself as quite accurate for propagating low-dimensional systems even with pronounced anharmonicity.
However, due to the well-known sign problem which is caused by the rapid oscillations and the divergence of classical monodromy matrices, convergence for systems of increasing dimensionality can be hardly achieved.
Hence, further approximations are performed in order to improve the performance for larger systems.

As a first approximation, the \LSC\ method is introduced, which is based on linearizing the \HK\ formula with respect to the initial displacement $\Delta z_0$ of the two trajectories~\cite{Miller-JCP-1998}.
This amounts to a fully classical propagation launched from a quantum-mechanically correct initial state $W_0(z)$
\begin{equation}
\label{eq: LSC}
W^{}_{\mathrm{\LSC}}(z,t) = \intop \diff z_0 \, \delta(z-z_t) W_0(z_0)
\enspace ,
\end{equation}
where the delta-function enforces the observables to be evaluated at time $t$.
While quantum effects present in the initial state are taken into account, those that develop during the dynamics are completely ignored, although it should be noted that the classical propagation is exact for treating harmonic oscillators.

Another approximation to the \HK\ propagator, particularly tailored for Gaussian initial states, can be obtained by expanding all exponents of the integrands in the \HK\ formula up to second order~\cite{Grossmann-JCP-2006}.
The resulting Gaussian integral over initial values can be then performed analytically, leading to the so-called \TGWD.
In Wigner representation, one obtains a Gaussian Wigner function for the time-evolved wave packet that is localized on a single classical trajectory, i.e.
\begin{equation}
\label{eq: TGWD}
W^{}_{\mathrm{TGWD}}(z,t)=\mathcal N_t \exp \left [ -\frac{1}{2 \hbar} (z-z_t)^T \Gamma_t (z-z_t) \right ]
\enspace ,
\end{equation}
with a proper normalization constant $\mathcal N_t$.
The positive-definite $(2f \times 2f)$ matrix $\Gamma_t$ is given in terms of the monodromy matrix $\mathbf M_t$ of the classical trajectory as
\begin{equation}
\Gamma_t= ( \mathbf M_t^{-1})^ T  \Gamma_0 \mathbf M_t^{-1} \enspace ,
\end{equation}
with the matrix $\Gamma_0$ of the initial Gaussian Wigner function.
The main simplicity of the \TGWD\ is that only a single trajectory starting from the center of the initial Gaussian Wigner function is needed instead of a full trajectory ensemble. 
This makes the \TGWD\ expression in \Eq{eq: TGWD} applicable even to quite complex systems.
However, this tremendous simplification comes at a price: \Eq{eq: TGWD} implies that negativities, and thus quantum coherences, can not be properly accounted for, as the resulting Gaussian Wigner function is strictly positive.
Again, the \TGWD\ yields exact results if a harmonic oscillator is propagated starting from a Gaussian initial state.

The last class of methods used in this paper constitutes hybrid schemes.
%
Having in mind that both the \LSC\ and \TGWD\ methods can be very inaccurate in some cases, an appealing idea is to maintain parts of the original full \HK\ formula.
One can, therefore, follow the idea of a system-bath partitioning and declare a small subset of \DOFs\ to be important and the remaining ones as unimportant.
The important \DOFs, denoted as $z_\mathrm{hk}$, may be then treated accurately on the level of the full \HK\ formula, whereas the rest is treated on a lower level using \TGWD\ or \LSC.
Since the hybrid schemes lead to very cumbersome expressions, we refrain from elaborating on all details but rather discuss their general structure.
Upon hybridizing the \HK\ propagator with the \LSC, see Ref.~\cite{Koda-JCP-2016}, one finds for the reduced Wigner functions in terms of the \HK\ variables $z_\mathrm{hk}$
\begin{eqnarray}
\label{eq: HK-LSC}
W^{}_{\mathrm{HK-LSC}}(z_\mathrm{hk},t) & = & \displaystyle \intop \diff z_{0,\mathrm{lsc}} \intop \frac{\diff \bar z_{0,\mathrm{hk}} \, \diff \Delta z_{0,\mathrm{hk}}}{(2 \pi \hbar)^{2f_\mathrm{hk}} } \, \, \tilde \mathcal C_t(\bar z_0, \Delta z_{0,\mathrm{hk}}) \, \e^{\frac{\i}{\hbar} \tilde S_t(\bar z_0,\Delta z_{0,\mathrm{hk}})}  g(z_\mathrm{hk};\bar z_{t,\mathrm{hk}}, \Delta z_{t,\mathrm{hk}}) \ \nonumber \\ 
& & \times  W_{0, \mathrm{lsc}}(z_{0,\mathrm{lsc}}) \intop \diff z_\mathrm{hk}' \, g^*(z_\mathrm{hk}'; \bar z_{0,\mathrm{hk}}, \Delta z_{0,\mathrm{hk}}) W_{0,\mathrm{hk}}( z_\mathrm{hk}') \enspace .
\end{eqnarray}
This expression resembles the \HK\ formula in terms of trajectory pairs for the $2 f_\mathrm{hk}$-dimensional subspace of the (relevant) \HK\ variables $z_\mathrm{hk}$.
The functions $g(z_\mathrm{hk};\bar z_{\mathrm{hk}}, \Delta z_{\mathrm{hk}})$ are, thus, defined in the same way as in \Eq{eq: coherent state phase space}, but limited to the variables $z_\mathrm{hk}$ only.
The less important \DOFs, $z_\mathrm{lsc}$, are treated on the level of the \LSC\ method for which initial midpoints and differences reduce to $\bar z_{0,\mathrm{lsc}} = z_{0,\mathrm{lsc}}$ and $\Delta z_{0,\mathrm{lsc}}=0$, i.e.\ the two trajectories are launched from the same initial point.
While the definition of the action $\tilde S_t(\bar z_0, \Delta z_{0,\mathrm{hk}})$ remains unchanged, the prefactor $\tilde \mathcal C_t(\bar z_0, \Delta z_{0,\mathrm{hk}})$ now takes
a more complicated block form in terms of the monodromy matrices, see Ref.~\cite{Koda-JCP-2016} for the precise definition.
Further, in full accordance with the standard treatment of quantum dissipative dynamics, it is assumed that the initial Wigner functions of the \LSC\ and \HK\ \DOFs\ are factorizable, i.e.\ $W_0(z)=W_{0,\mathrm{hk}}(z_\mathrm{hk})W_{0,\mathrm{lsc}}(z_\mathrm{lsc})$.

In a similar way, a hybrid expression combining the \HK\ formula with the \TGWD\ method can be formulated.
Carrying out the \TGWD\ approximation for a subset of \DOFs, $z_\mathrm{tg}$, only, the reduced Wigner function for the \HK\ \DOFs\ reads
\begin{eqnarray}
\label{eq: HK-TG}
W^{}_{\mathrm{HK-\TGWD}}(z_\mathrm{hk},t) & = & \frac{1}{\pi \hbar} \intop \frac{\diff \bar z_{0,\mathrm{hk}} \, \diff \Delta z_{0,\mathrm{hk}}}{(2 \pi \hbar)^{2f_\mathrm{hk}} (2\hbar)^{2f_\mathrm{tg}}} \sqrt{ \frac{\det \gamma \cdot R (R')^*}{\det A \cdot \det H} } \sqrt{\frac{1}{-\Lambda_{11} + 2 \Lambda_{12} - \Lambda_{22}}} \nonumber \\
& & \times \exp \left [ \tilde e + h + \frac{\i}{\hbar} (S^+ - S^-) \right ] \braket{g_{\gamma_S}| \Psi_\alpha} \braket{\Psi_\alpha | g_{\gamma_S} } \enspace .
\end{eqnarray}
The ingredients of this cumbersome expression are not reiterated here since they are given in Ref.~\cite{Buchholz2012}.
However, the general structure of the reduced Wigner function is similar to the hybrid expression in \Eq{eq: HK-LSC}.
The integration over initial values of the \HK\ \DOFs\ $z_\mathrm{hk}$ is maintained and, the counterparts of the action $\tilde S_t$ and the \HK\ prefactor $\tilde C_t$ are hidden in the quantities $S^\pm$ and $R$, see Ref.~\cite{Buchholz2012}.
The variables $z_\mathrm{tg}$ are, in turn, always propagated from the centers of the initial Gaussian wave packet and, hence, are not integrated over, which manifests the single-trajectory character of the \TGWD\ method.

\section{Numerical Tests}
\label{sec: numerical tests}
The two versions of the Fourier method are now applied to calculating spectral densities from explicit semiclassical simulations of a system and its environment.
Although it would be interesting to consider realistic systems in condensed phase, a previous study has shown that due to the ``invertibility problem'' a mapping onto the CL model can be inconsistent and, thus, the resulting spectral density is unphysical~\cite{Gottwald-JPCL-2015}.
In order to switch off this problem from the outset, a \CL\ model bath is imposed in the following.
Here, the intrinsic semiclassical errors of the aforementioned methods, i.e.\ the deviations of the statistically converged semiclassical averages from the exact quantum ones, are the only sources of errors and it can be judged on how strongly the spectral densities are affected by them.
For this purpose, spectral densities from semiclassical correlation functions and expectation values are compared to the model spectral densities that were imposed.

\subsection{Model Systems and Technical Details
	\label{sec: technical details}}
The considered one-dimensional systems
were described by a coordinate $q$ with mass $m=1$ and $\hbar=1$.
%
Three prototypical system potentials $\VS(q)$ were employed: i) 
a harmonic potential
\begin{equation}
\VS(q)=\frac{1}{2} \omega_0^2 q^2 \enspace ,
\end{equation}
with the harmonic frequency $\omega_0$; ii) an anharmonic Morse potential
\begin{equation}
\label{eq: Morse potential}
\VS(q)=D\cdot \left ( 1 - \e^{-\alpha q}\right )^2 \enspace ,
\end{equation}
with the dissociation energy $D$ and stiffness $\alpha$;  and iii) a quartic double-well potential 
\begin{equation}
\label{eq: double-well potential}
\VS(q)=-\frac{1}{4} q^2 + \frac{1}{64 E_b} q^4\enspace ,
\end{equation}
with the barrier height $E_b$.
%
A
pure state $\rho_\mathrm S=\ket{\psi_\mathrm S}\bra{\psi_\mathrm S}$, with the Gaussian wave function
\begin{equation}
\label{eq: initial wave function}
\braket{q | \psi_\mathrm S}=\left (\frac{\gamma_\mathrm S}{\pi} \right)^{-1/4} \exp \left [ -\frac{\gamma_\mathrm S}{2} (q-q_0)^2 \right ]
\end{equation}
was chosen as the initial state of the system.
%
In this case,
the static averages in \Eq{eq: xi from TCF} and \Eq{eq: xi from averages} take the values of $\langle pq \rangle = -\i/2$ and $\langle p^2 \rangle = \gamma_\mathrm S/2$  and $\langle q \rangle = q_0$ and $\langle p \rangle = 0$, respectively.

For the environment, a model spectral density composed of (superpositions of) Gaussian functions
%
\begin{equation}
J(\omega)=\eta \, \omega \, \exp \left [ -\Delta (\omega-\omega_\mathrm{C})^2 \right ] \enspace 
\end{equation}
was chosen.
%
%
The parameter $\omega_\mathrm{C}$ was always tuned close to a characteristic system frequency such that a resonant energy transfer between the system and bath occurs.
In all simulations, an explicit representation of the bath in terms of $N_\mathrm{osc}=20$ - $40$ oscillators, coupled to the system according to \Eq{eq: CL potential}, was used.
The frequencies $\omega_i$ of the bath oscillators were set equidistantly in an interval $[\omega_s; \omega_e]$, i.e.
\begin{equation}
\omega_i = \omega_s + i\cdot \Delta \omega \enspace \enspace \enspace i=0,...,N_\mathrm{osc} -1
\end{equation}
with $\Delta \omega =(\omega_e - \omega_s)/N_\mathrm{osc}$.
The coupling strengths $g_i$ were set according to the model spectral density $J(\omega)$ as
\begin{equation}
g_i=\sqrt{\frac{2}{\pi} \, J(\omega_i) \, \omega_i \, \Delta \omega} \enspace .
\end{equation}
The overall initial state of system and bath was factorized~$\rho = \rho_\mathrm S \cdot \rho_\mathrm B$ with the bath density matrix given by \Eq{eq: bath density} according to different temperatures $T$, see results section below.
All specific parameter choices are given in the results section for the different cases considered. 

For each setup, the correlation functions $C_{pp}(t)$ and $C_{pF}(t)$ and the expectation values $\langle p(t) \rangle $ and $\langle F(t) \rangle$ were computed in order to check how well the model spectral density is reproduced 
when these quantities are inserted into \Eq{eq: xi from TCF} and \Eq{eq: xi from averages}, respectively.
This test was performed for the semiclassical techniques described in \Sec{sec: semiclassics}, i.e.\ the \LSC\ method, the \TGWD\ method and the hybrid schemes, combining the \HK\ formula with \LSC\ (\HK-LSC) and with \TGWD\ (\HK-\TGWD).
When using the two hybrid schemes, the full \HK\ formula was always maintained for the system coordinate $q$, whereas the bath oscillators were treated via the corresponding lower accuracy method.
Additionally, full quantum simulations using the \MCTDH\ method were run using the Heidelberg package~\cite{MCTDH} in order to give access to the exact quantum expectation values $\langle p(t) \rangle $ and $\langle F(t) \rangle$ for the zero-temperature regime 
(see results section).
Semiclassical quantities were evaluated via importance sampling using $10^5-10^7$ classical trajectories, except for the full \TGWD\ approach, where just a single trajectory started from the center of the initial wave packet with zero momentum was employed.
The full expressions for the estimators and sampling densities of all considered quantities are given in the supplement.
For the classical propagation of trajectories and monodromy matrices the Velocity-Verlet algorithm was used with a time step of $\Delta t= 0.1$ (harmonic and Morse oscillators) or $\Delta t=0.05$ (double-well potential) and semiclassical quantities were computed every 5th time step.

As it was intensively discussed in Refs.~\cite{Gottwald-JPCL-2015,Gottwald-JCP-2016}, the numerical noise stemming from the usually unconverged tails of the time domain functions needed to be properly reduced.
For this purpose, a Gaussian low-pass filter was applied before the half-sided Fourier transform, i.e.\ the time-domain functions were multiplied by a Gaussian window $G(t)=\exp [ -t^2/(2 T^2)]$ with the window width $T=70$ (harmonic and Morse oscillators) or $T=100$ (double-well potential), which roughly corresponded to the characteristic decay time.
This procedure also suppressed artificial revivals after a characteristic time $2\pi/\Delta \omega$, which stem from the discretization of the bath in intervals of the length $\Delta \omega$.
As was discovered in Ref.~\cite{Gottwald-JCP-2016}, the Fourier method has a phase-sensitive error accumulation behavior, which can critically depend on the noise reduction scheme employed.
Thus, before non-linear dynamics was considered, the impact of smoothing errors from the Gaussian filtering had been analysed in the harmonic regime following the same lines as in Ref.~\cite{Gottwald-JCP-2016}.
As shown in detail in the supplement, the error $\hat \epsilon_\xi (\omega)$ of the memory kernel can be expressed as
\begin{eqnarray}
\label{eq: error formula}
| \hat \epsilon_\xi (\omega)| & = & \Bigg |  \frac{\hat C_{pF}\num(\omega) }{\hat C_{pp}\num (\omega) + m \langle pq \rangle } \Bigg | \cdot \Big |  r_{C_{pF}}(\omega) - r_{C_{pp}}(\omega) \e^{\i \Delta \phi (\omega)} \Big | \nonumber \\
| \hat \epsilon_\xi (\omega) | & = & \Bigg |  \frac{\langle \hat F(\omega)  \rangle \num }{\langle \hat p(\omega) \rangle \num + m \langle q \rangle } \Bigg | \cdot \Big |  r_{F}(\omega) - r_{p}(\omega) \e^{\i \Delta \phi(\omega)} \Big | \enspace ,
\end{eqnarray}
where the superscript "$(\mathrm{num})$" denotes the numerically obtained dynamical quantities, while $r(\omega)$ and $\Delta \phi$ stand for the absolute value and the phase difference of their complex relative errors, respectively.
These equations imply that the phase difference $\Delta \phi(\omega)$ decides on error accumulation or cancellation.
If the absolute values of the errors $r(\omega)$ are similar for both quantities and if the phase difference is close to a multiple of $2 \pi$, the error will cancel.
In contrast, if the phase difference is close to an odd multiple of $\pi$, the error will accumulate.

\subsection{Error Analysis in the Harmonic Regime}
\label{sec: harmonic regime}
First, a harmonic oscillator with the frequency $\omega_0=0.62$, adopted from an earlier semiclassical study of dissipative dynamics~\cite{Koch-CP-2010}, and an initial Gaussian wave packet with $\gamma_\mathrm S=1$ and $q_0=5$ is considered.
The bath is described via a spectral density consisting of a single Gaussian function with parameters $\eta = 0.1$, $\Delta = 80$ and $\omega_\mathrm{C} = \omega_0$, mimicked by 20 bath oscillators placed equidistantly in the frequency interval $[\omega_s; \omega_e]=[0.3;0.9]$.
Since temperature does not play any role in the harmonic regime, only the zero-temperature case is considered here.
%
Importantly, the exact results for $\hat C_{pp}(\omega)$ and $\hat C_{pF}(\omega)$
as well as
$\langle \hat p(\omega) \rangle$ and $\langle \hat F(\omega) \rangle$ 
can be obtained analytically (see supplement) and, further, all semiclassical methods are numerically exact in the harmonic regime.
Thus, the impact of purely numerical errors, especially the smoothing errors from Gaussian filtering, can be estimated and analyzed using
the error formula
\Eq{eq: error formula}.
In order to obtain statistically converged results, $10^5$ trajectory pairs for each semiclassical scheme are employed, apart from the pure \TGWD\ method which is based on a single trajectory only.

\begin{figure}[htb]
\includegraphics[width=0.99\textwidth]{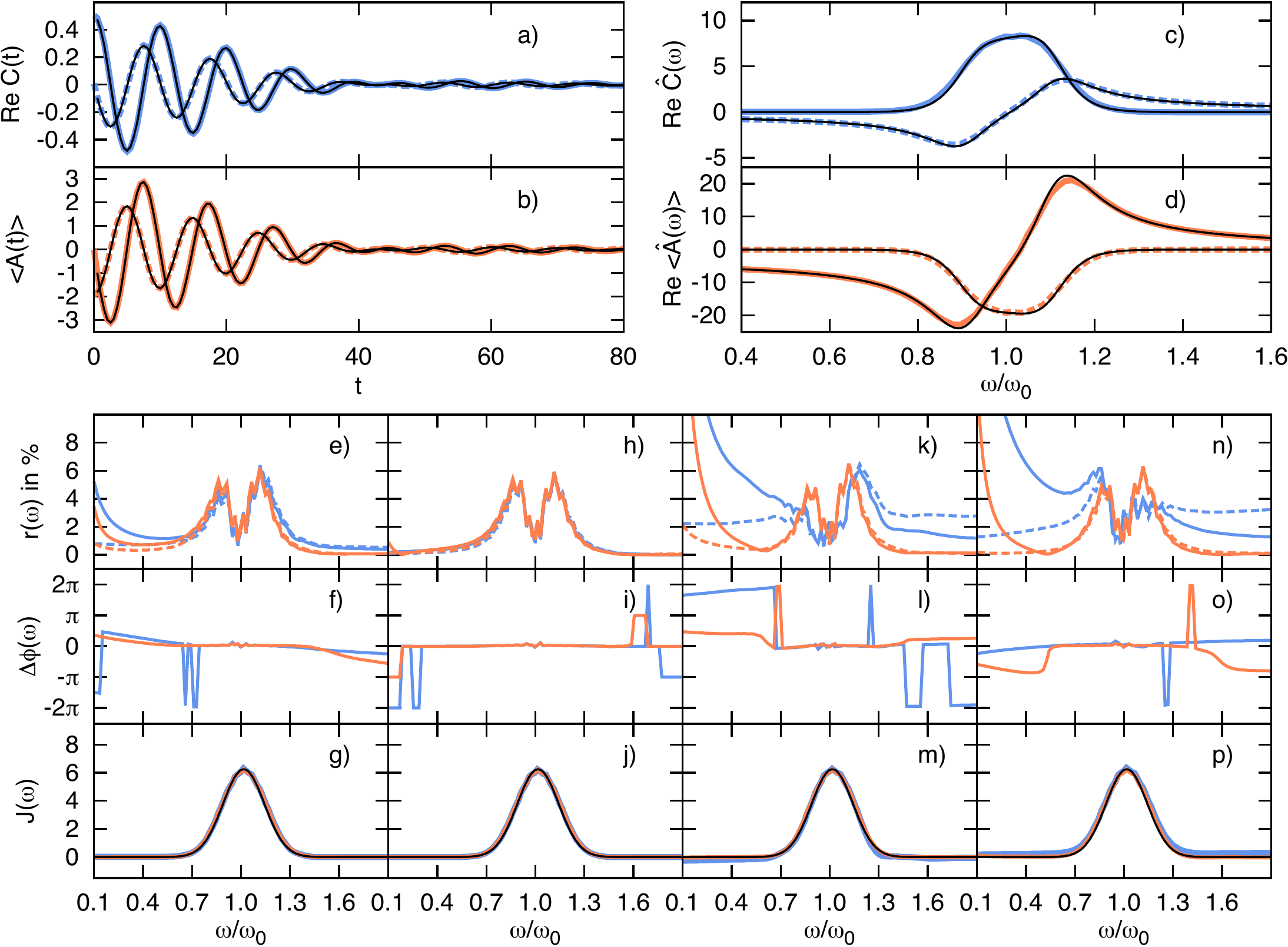}
\caption{
\label{fig: error-harmonic}
The error analysis for the two computational protocols based on i) correlation functions (blue curves) and ii) expectation values (orange curves).
Quantities involving system forces are represented by dashed lines, those involving momenta only by solid lines. 
Panels a) - d) display time and frequency domain results for correlation functions, labeled $C(t)$ for the \MAF\ and \MFC, and expectation values, labeled $\langle A(t) \rangle$ for $p$ and $F$, obtained from the hybrid \HK-\TGWD\ method together with the exact curves (black).
Below, the absolute values and phase differences of the relative errors as well as the resulting spectral densities are shown for the \LSC\ method (panels e - g), the \TGWD\ method (panels h - j), the \HK-LSC hybrid method (panels k - m) and the \HK -\TGWD\  hybrid method (panels n - p) each for the two computational protocols.
Exact spectral densities are displayed in black color therein.
}
\end{figure}

In \Fig{fig: error-harmonic}, all semiclassical ingredients of the error formula are shown for the two protocols, i.e.\ \Eq{eq: xi from TCF} using correlation functions and \Eq{eq: xi from averages} based on expectation values.
In the upper part, panels a) - d), the semiclassical correlation functions and expectation values are shown in frequency and time domains for the \HK-\TGWD\ hybrid scheme only. 
The results from all other methods look exactly the same.
Comparing all quantities to the exact references (black lines) reveals a good accuracy for the chosen numerical setup.
This is further underlined by small relative errors whose absolute values are displayed in the third row  of \Fig{fig: error-harmonic}.
For all semiclassical methods
considered, the error magnitudes are similar and below $8\%$ in the resonant region around the system frequency, i.e.\ $\omega/\omega_0=1$.
The phase differences, shown in the fourth row, are always close to zero in the resonant frequency region, whereas there are phase jumps being mostly a multiple of $2  \pi$ outside.
Thus, the favorable regime of error cancellation is obtained resulting in very accurate spectral densities, shown in the fifth row therein.
In the off-resonant regions, however, the hybrid schemes yield phase differences sometimes close to $\pm \pi$, see e.g.~panel o), and increasing error magnitudes in the lower frequency region, see panels k) and n).
However, this has only little impact on the resulting spectral densities, which are slightly above zero in these regions, see panels m) and p).
All in all, the chosen numerical setup yields a reasonable accuracy of the spectral densities for all semiclassical methods considered 
in combination with
the two protocols based on correlation functions and expectation values.
It is confirmed again, that the Gaussian filtering scheme is the method of choice for a reasonable noise reduction being fully compatible with the phase-sensitive error cancellation.

\subsection{Anharmonic Regime at Zero Temperature}
\label{sec: anharmonic zero temperature}
%
Having verified 
that the window width used in Gaussian filtering is
set up reasonably, we now pay attention to the more interesting anharmonic regime for which the semiclassical propagators are not exact anymore and semiclassical errors can occur.
Here, a Morse oscillator was considered, with the parameters $D=30$ and $\alpha=0.08$ leading to the harmonic frequency $\omega_0=0.62$, i.e.\ the same as in the harmonic case analyzed above.
The same initial state was also adopted and probes a strong anharmonicity region in the Morse potential.
Two model spectral densities were employed, with the first one coinciding with the one used for the harmonic case.
The second spectral density was a superposition of two Gaussian functions with the parameters $\eta_1=\eta_2 = 0.1$, $\Delta_1 = 100$, $\Delta_2=200$, $\omega_{\mathrm{C},1} = 0.55$ and $\omega_{\mathrm{C},2}=0.7$ sampled by 40 bath oscillators in the frequency interval $[0.3;0.9]$.
The latter setup was chosen in order to test whether semiclassical methods are sensitive
to the presence of structure in the
spectral density.
Again, $10^5$ trajectories were needed for statistical convergence, apart from the single-trajectory \TGWD\ scheme.
Here, we investigate the zero-temperature regime, for which exact quantum-mechanical references can be obtained using the \MCTDH\ method, see supplement for technical details.
The investigations are extended to the finite-temperature case in the next subsection.

\begin{figure}[tb]
\includegraphics[width=0.99\textwidth]{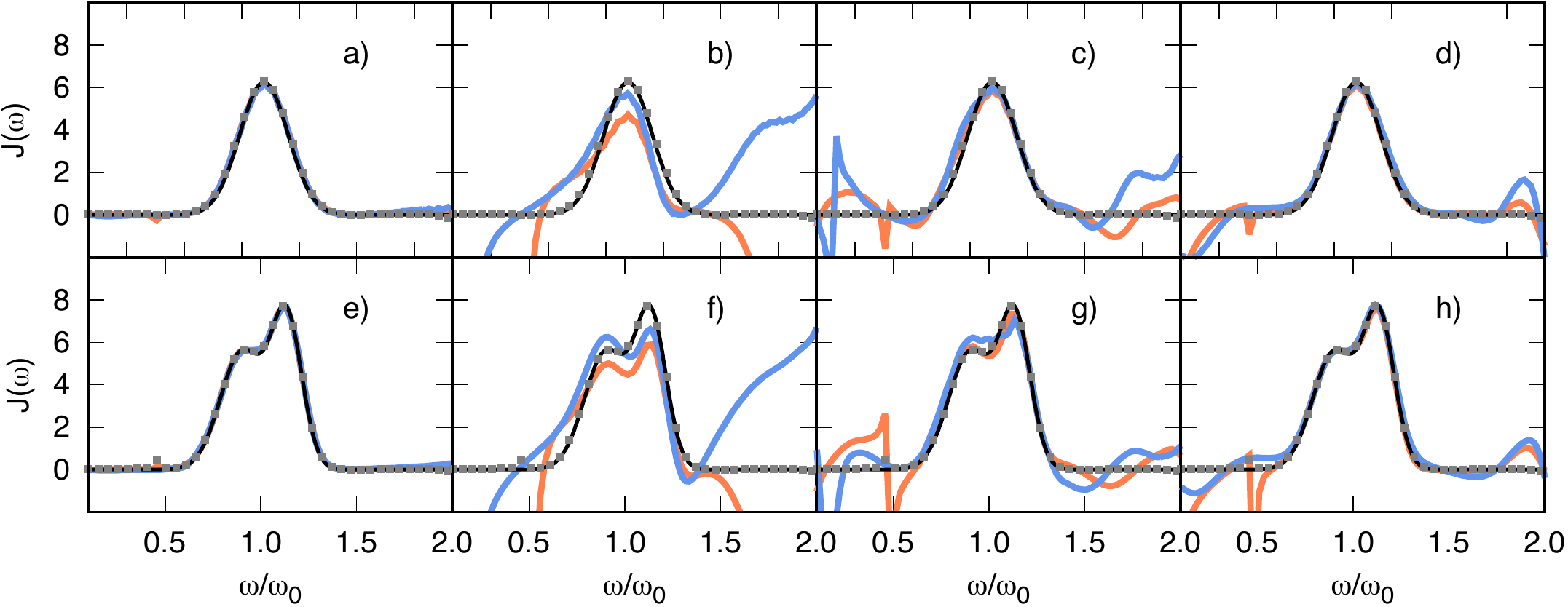}
\caption{
\label{fig: anharmonic}
Spectral densities resulting from the two protocols based on correlation functions (blue curves) and on expectation values (orange curves) are shown for the \LSC\ method (panels a and e), the \TGWD\ method (panels b and f), the hybrid \HK-LSC approach (panels c and g) and the hybrid \HK -\TGWD\ ansatz (panels d and h).
The analytic curves and the results from \MCTDH\ using the Fourier method based on expectation values are added in each panel in black curves and grey dots.
In the upper row, a single Gaussian model spectral density has been imposed whereas in the lower row a superposition of two Gaussian functions has been employed, see text for parameters. 
}
\end{figure}
In \Fig{fig: anharmonic}, the resulting spectral densities are shown for all semiclassical methods
considered combined with the two computational protocols based on 
expectation values and correlation functions.
In each panel, the analytic spectral density (black curves) as well as the one obtained from the
numerically exact \MCTDH\ method (grey dots) are added.
The \MCTDH\ curves are always in perfect agreement with the exact model spectral densities.
Since \MCTDH\ is a numerically exact exact propagation scheme, this confirms that the proposed Fourier method is a conceptually correct way to obtain spectral densities from quantum-dynamical simulations.
Starting the analysis of approximate approaches from the low-level ones, 
one observes that the fully classical \LSC\ scheme yields spectral densities that perfectly coincide with the analytic and \MCTDH\ curves, see panels a) and e).
In contrast, the single-trajectory \TGWD\ method, panels b) and f), yields rather poor results in the resonant region overestimating the width and underestimating the magnitudes.
Further, large artefacts develop in the off-resonant region, which are even comparable in magnitude to the actual spectral density in the resonant region.
Turning to the hybrid schemes, the \HK-LSC method yields reasonable results in the resonant region for both spectral density types considered, see panel c) and g), although the single Gaussian spectral density, panel c), is slightly broadened.
At higher and lower frequencies additional artefacts emerge, but these are much smaller in their magnitude than that for the \TGWD\ method.
A similar behavior is observed for the \HK-\TGWD\ method, but its accuracy is overall better, especially in the resonant region.
%

The fact that the \LSC\ simulations yield perfect spectral densities is not very surprising since the system is nearly classical, see supplement.
%
%
%
For the other methods, deducing the reasons for the observed artefacts is rather difficult.
Since the errors due to the parametrization of spectral densities were shown to behave well and can be thus eliminated, as well as the statistical errors of the simulation, 
we conclude that the artefacts must be due to the intrinsic semiclassical errors of the methods.
Fortunately for the hybrid methods, these artefacts occur mostly in the off-resonant region and are, thus, generally not problematic, as the vibrational dynamics is usually not sensitive to them.

\subsection{Anharmonic Regime at a Finite Temperature}
After having obtained the evidence that the hybrid schemes yield reasonable spectral densities in the resonant region at zero temperature, an impact of a finite temperature on them is investigated.
Thereby the same Morse oscillator setup is employed and the spectral densities obtained at a temperature $kT=1$ are compared to that at zero temperature discussed above.
Since a generic \CL\ model bath is used, the spectral density must be strictly independent on temperature, see \Eq{eq: spectral density}, which provides a sensitive measure of self-consistency.

\begin{figure}[htb]
\includegraphics[width=0.99\textwidth]{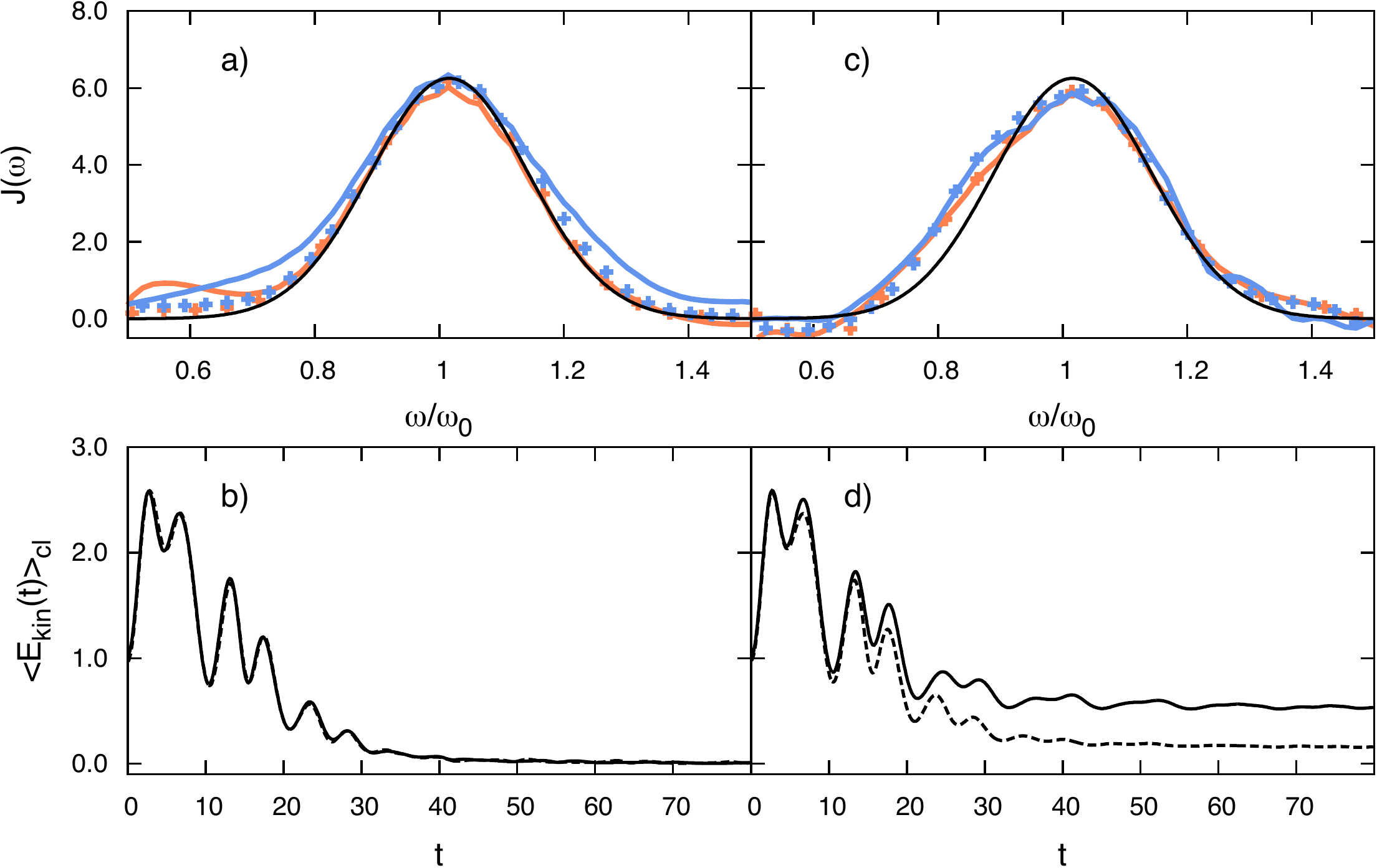}
\caption{
\label{fig: anharmonic-beta-1}
Semiclassical results for the single-Gaussian spectral density at finite temperature $kT=1$ are presented for the hybrid schemes \HK-\TGWD\ (panel a) and \HK-LSC (panel c).
Orange and blue curves stand for the computation protocol based on expectation values and on correlation functions, respectively, black curves denote the exact spectral density.
For a better comparison, the zero-temperature results are replotted with points.
In panels b) and d), the corresponding evolution of the averaged kinetic energy of the underlying classical propagation is shown. Solid curves stand for $kT=1$ and dashed ones for $kT=0$.
}
\end{figure}

In the upper row of \Fig{fig: anharmonic-beta-1}, the spectral densities (solid curves) from the hybrid methods are shown for $kT=1$ together with the exact curves (black) and the zero-temperature results from \Fig{fig: anharmonic} (replotted with points).
It turns out that the spectral densities from the \HK-LSC method, panel c)
are of similar quality as in the zero-temperature case for both computation protocols.
A different trend is seen, however, for the \HK-\TGWD\ method,
see panel a) therein.
While for the computation protocol using expectation values the results are correct in shape and magnitude, apart from deviations starting below $\omega/\omega_0 =0.7$, the spectral density resulting from correlation functions is broadened compared to its zero-temperature counterpart.
%
Thus, semiclassical errors 
might be
more sensitive to the temperature for the \HK-\TGWD\ method than they are for the \HK-LSC propagator.

Although analyzing and controlling the intrinsic semiclassical errors of the two hybrid methods is difficult,
a possible explanation for the observed behavior can be found in the way how temperature is incorporated into the two methods.
In the \HK-\TGWD\ method, the \IVR\ integral, see \Eq{eq: HK-TG}, is taken with respect to the system \DOFs\ only.
Consequently, each bath trajectory is started from the fixed values $Q_i=0$ and $P_i=0$, which always corresponds to the classical zero-temperature case.
The true temperature is accounted for via Boltzmann factors only, which are hidden in the variables $A$, $H$, $\tilde e$ and $h$ of \Eq{eq: HK-TG}, see Refs.~\cite{Buchholz2012, Goletz2009}.
Contrary to this, the \HK-LSC method employs an average over bath \DOFs\ according to the quantum-mechanically correct thermal Wigner function, see \Eq{eq: HK-LSC}.  
Thus, true temperature fluctuations are reflected in the corresponding classical propagation on which the semiclassical expression is based.
The consequences of these different treatments are shown in \Fig{fig: anharmonic-beta-1} for the \HK-\TGWD\ method (panel b) and the \HK-LSC approach (panel d).
Here, the classically-averaged kinetic energies of the Morse oscillator are shown for $kT=0$ (dashed lines) and $kT=1$ (solid lines).
For the \HK-\TGWD\ scheme, one observes that the kinetic energy is fully dissipated into a bath since it effectively acts at zero temperature, irrespectively of the true temperature that is intended.
Hence, the semiclassical \HK-\TGWD\ averages are always drawn from an unphysical zero-temperature dynamics and it is questionable whether this method is able to provide a correct temperature behavior at all.
This might well be the reason, why the corresponding spectral densities are reproduced less accurately with increasing temperature.
In contrast, the averaged kinetic energies for the \HK-LSC method stabilize at non-zero values, reflecting
qualitatively the zero-point energy and temperature fluctuations.
Thus, the quality of the spectral densities is unaffected upon increasing the temperature and the observed discrepancies are more likely to originate from other semiclassical errors.
It should be noted, however, that the accuracy of the spectral density is still best for the \HK-\TGWD\ approach if the expectation-values protocol is used.

\subsection{Dynamical Quantum Effects}
The aforementioned studies of the Morse oscillator indicate that the semiclassical hybrid schemes can be promising tools for computing spectral densities in the presence of quantum effects.
However, although a strong anharmonicity is probed in this parameter setup, no quantum effects were visible in the considered dynamical quantities.
Hence, a natural question arises how the semiclassical hybrid methods perform in a regime, 
where the quantum effects are strong and visibly manifest themselves in the observables.
For this purpose, the Morse parameters are modified to $D=3$ and $\alpha=0.253$, which increases the anharmonicity further but keeps the harmonic frequency of $\omega_0=0.62$ unchanged.
The Morse oscillator is coupled to the same bath with the single-Gaussian spectral density,
i.e.\ $\eta=0.1$, $\Delta=80$ and $\omega_\mathrm{C}=\omega_0=0.62$.
In order to avoid dissociating trajectories, the initial displacement of the wave packet is reduced to $q_0=1$.
As a second example, we employ the double-well potential in \Eq{eq: double-well potential} with a barrier height of $E_b=0.5$.
In this setup, the ground state energy is very close to the barrier top, whereas the first excited state is already well above.
As an initial state, we choose a Gaussian wave packet centered at $q_0=-2.5$ and $\gamma_\mathrm S=1.379$, which yields an energy of 0.35 above the barrier.
Hence, the deep-tunneling regime, which is known to be problematic for real-time semiclassical propagation, is excluded from consideration here.
The bath is described by a single-Gaussian spectral density with parameters $\eta=0.1$, $\Delta=80$ and $\omega_\mathrm{C}=0.68$ (close to the 0-2 transition frequency) and is represented by 20 oscillators in the frequency interval $[0.4;0.9]$.
In order to achieve convergence for the \HK-LSC method, $10^7$ trajectories for the Morse oscillator, whereas the double-well potential could not be converged due to the infamous sign problem.
For the other methods, however, $10^5$ trajectories were still sufficient.

\begin{figure}[tb]
\includegraphics[width=0.99\textwidth]{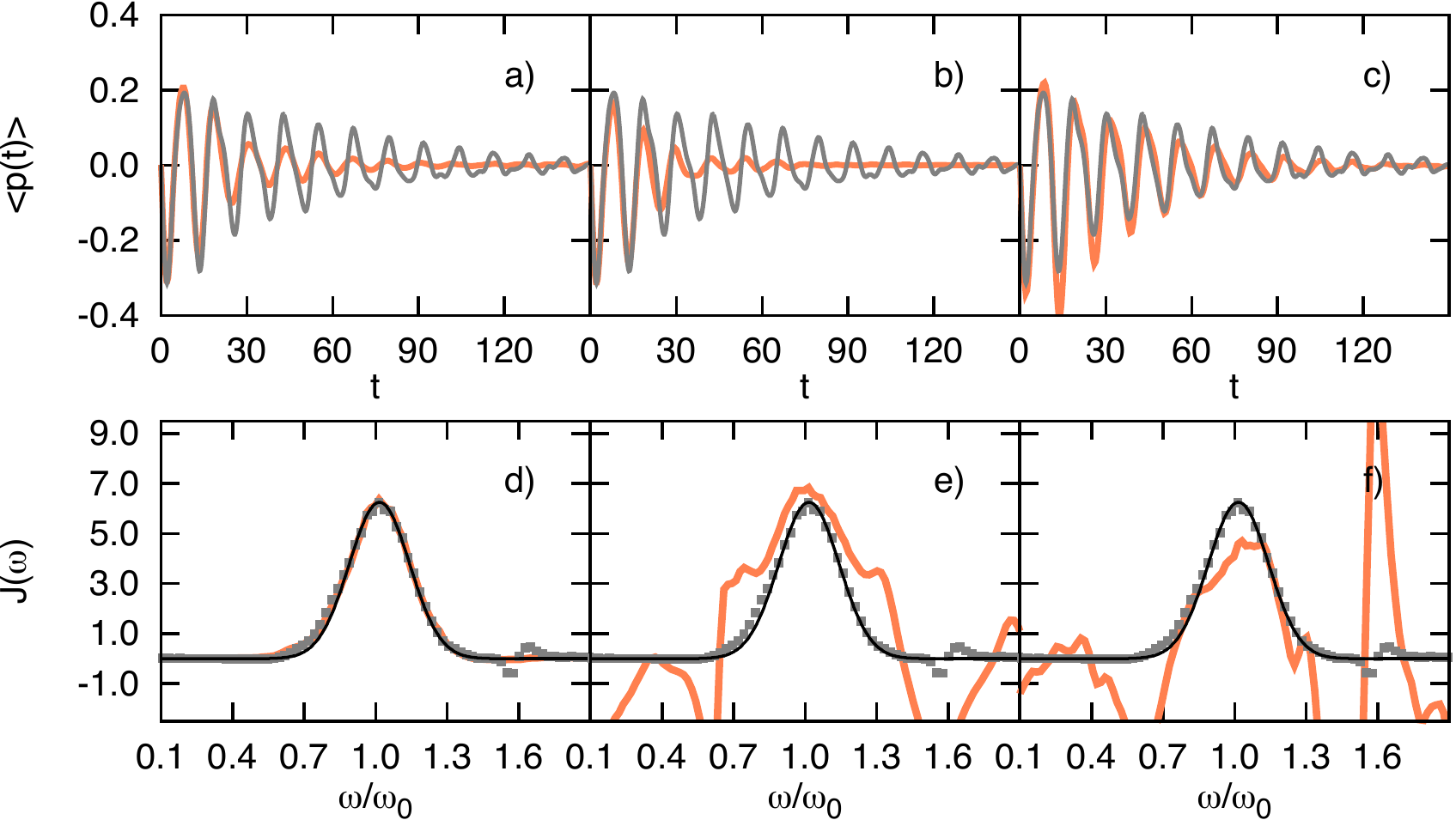}
\caption{
\label{fig: Morse quantum}
Momentum expectation values (upper row) of the Morse oscillator and corresponding spectral densities (lower row) are shown for the semiclassical (orange curves) and exact \MCTDH\ simulations 
(grey dotted line).
Panels a), d) correspond to the \LSC, panels b), e) to the \HK-\TGWD\ and panels c), f) to the \HK-LSC methods.  Black curves in the lower row indicate the exact model spectral density employed.}
\end{figure}
In \Fig{fig: Morse quantum}, the momentum expectation values of the \LSC\ and semiclassical hybrid methods (orange curves) are shown together with the exact \MCTDH\ results depicted in gray.
Looking at panel a), one observes that the classical \LSC\ curve reveals a stronger damping as \MCTDH\ one.
This finding can be explained by the fact that the 0-1 transition frequency of the quantum Morse oscillator is 
red-shifted compared to the harmonic frequency and is, thus, slightly off-resonant to the bath.
Nonetheless, the spectral densities (panel d) coincide perfectly as was already observed in \Sec{sec: anharmonic zero temperature}.
For the \HK-\TGWD\ hybrid scheme (panels b and e), one observes an overestimated damping of the momentum expectation value and, hence, the characteristic quantum feature is not reproduced at all.
These discrepancies accumulate to large errors in the corresponding spectral density, even in the resonant frequency region.
Although the \HK-LSC result for the expectation value (panel c) qualitatively fits better to the quantum result, the quantitative deviations spoil the spectral density even stronger (panel f).
Thus, it becomes apparent that, unfortunately, the semiclassical hybrid methods are not able to account for stronger dynamical quantum effects and, as a consequence, yield completely useless spectral densities.
\begin{figure}[tb]
\includegraphics[width=0.99\textwidth]{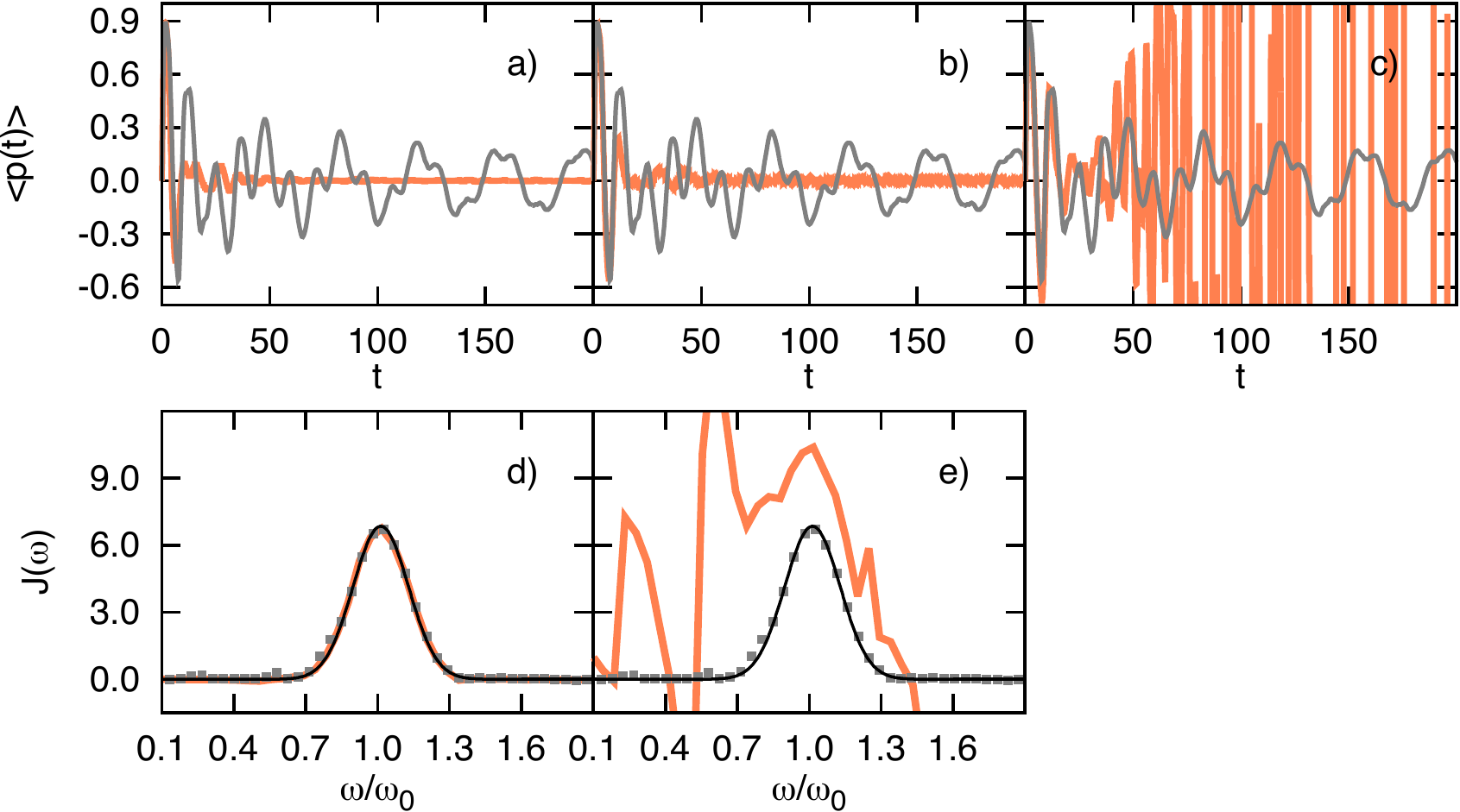}
\caption{
\label{fig: double well}
Momentum expectation values and corresponding spectral densities for the double well potential. The layout is the same as in \Fig{fig: Morse quantum}. For the \HK-LSC method, no converged expectation values and, hence, no spectral density could be obtained.}
\end{figure}

The same trend is observed for a particle in the double-well potential, see \Fig{fig: double well}.
While the momentum expectation value from \LSC\ is quickly dampened to zero, the quantum curve possesses a long-living low frequency oscillation which stems from the 0-1 transition, describing a shallow tunneling oscillation between the two wells.
Although this striking quantum effect is absent in the \LSC\ result, the classical and quantum spectral densities coincide perfectly (panel d).
Again, the \HK-\TGWD\ method (panel b) yields neither a quantitatively nor a qualitatively correct description of momentum expectation value.
The \HK-LSC results (panel c) strongly suffer from the sign problem, as it could not be statistically  converged even with $10^7$ trajectories but, given the trend observed in \Fig{fig: Morse quantum}, a good quantitative agreement is not to be expected in any case.
%
%
As a consequence of the large discrepancies, the spectral density is, again, not at all reproducible, see panel e) for the \HK-\TGWD\ approach.

These investigations imply that a fully classical simulation is sufficient for spectral density calculations if the bath can be properly described by the \CL\ model.
Surprisingly, this is even true if strong quantum effects show up in the observable under study and if the \LSC\ method does not give a quantitatively correct description. 
The explanation of this remarkable result is that the underlying \GLE\ is valid in both, the purely classical and fully quantum, regimes employing the very same definition of the spectral density as a coupling-weighted frequency distribution.
The hybrid schemes are meant to be an approximation to the quantum dynamics but, in contrast, fulfill neither a classical nor a quantum \GLE\ strictly.
Thus, they are incapable of reproducing the correct spectral densities in the presence of strong dynamical quantum effects.
Although, the qualitative agreement between the \MCTDH\ results and the \HK-LSC method is reasonable, the intrinsic semiclassical errors lead to completely wrong spectral densities, which might be due to the critical error accumulation behavior of the proposed Fourier method.
However, the semiclassical errors are hard to control and we, thus, do not expect that the quality of the results can be systematically improved.

\section{Conclusions and Outlook}
\label{sec: conclusions}
In this paper, we have addressed the question of computing spectral densities for reduced quantum dynamics on the basis of the Caldeira-Leggett (\CL) model.
The main focus was on the situation in which quantum effects of the system dynamics can be taken into account explicitly.
For this purpose, we have reformulated the Fourier method, developed in Refs.~\cite{Gottwald-JCP-2015} and~\cite{Gottwald-JCP-2016} for the classical equilibrium case, according to the regime of quantum dissipation.
This reformulation has led to two possible computation protocols using, on one hand, the momentum-momentum and momentum-force correlation functions or, on the other hand, the momentum and force expectation values.
As an approximate way to compute the quantities, we have considered common semiclassical simulation techniques, which are the linearized semiclassical initial-value representation (\LSC), the thawed Gaussian wave packet dynamics (\TGWD) and the recently developed hybrid schemes combining the two with the more accurate Herman-Kluk (\HK) propagator, termed \HK-LSC and \HK-\TGWD, respectively.
In order to test the proposed concept, we have considered different systems coupled resonantly to a \CL\ model bath with a given model spectral density.
For this setup we have checked whether the imposed model spectral density is reproduced with sufficient accuracy by the semiclassical simulations.

While for a harmonic oscillator, all methods reproduced the exact spectral density with high accuracy, different levels of accuracy were obtained for an anharmonic Morse oscillator depending on the semiclassical method at hand.
The \LSC\ method, that is based on a fully classical propagation launched from the quantum-mechanically initial state, has turned out to yield perfectly accurate results, whereas the \TGWD\ method turned out inapplicable in general.
For the semiclassical hybrid schemes, a different behavior was observed depending on the presence of quantum effects.
If the Morse oscillator is considered close to the classical regime, the hybrid schemes turned out to be successful.
At zero temperature, both schemes reproduced the exact spectral density reasonably well in the important resonant region and no differences between the two computation protocols, i.e.\ from expectation values or correlation functions, could be observed.
At finite temperature, the \HK-\TGWD\ method yielded broadened spectral densities especially for the computation protocol from correlation functions.
In contrast, the quality of the results was temperature-independent for the \HK-LSC method.
We suppose that this behavior can be explained by the fact that in the \HK-\TGWD\ method, temperature fluctuations are not accounted for in the underlying classical propagation and, instead, the bath effectively acts at zero temperature always.
Nevertheless, it turned out that using the computation protocol from expectation values still yields accurate spectral densities in the resonant region.
Unfortunately, if the dynamical quantities show stronger quantum effects, as observed in a Morse potential with a decreased dissociation energy and in a quartic double well potential, both hybrid schemes yielded a quantitatively incorrect description and, thus, were unable to reproduce the imposed model spectral densities.
%
%
Very importantly, the purely classical \LSC\ simulation always yielded perfectly correct spectral densities, even if strong dynamical quantum effects, that were not at all reproduced, were present.
We have argued that the reason for this remarkable observation lies in the fact that the \GLE, on which the Fourier method is based, is strictly valid in both, the quantum and classical, regimes with exactly the same definition of the spectral density.
Hence, this underlines that using classical \MD\ simulations are completely sufficient for spectral density calculations, if the environment has the \CL\ model form.
In contrast, there is no need to employ a semiclassical approximation whose errors rather spoil the results significantly.

In a future study, we intend to repeat the investigations for more realistic environments for which the \CL\ model form is not assumed by construction.
For this situation, the developed Fourier method provides a way to map the true environment onto the \CL\ model resulting in an effective spectral density that tries to mimic its influence on the system by fluctuations and dissipation. 
Although we have shown in a previous study that such a mapping can be inconsistent due to the 'invertibility problem'~\cite{Gottwald-JPCL-2015}, chances are high to find a proper example in the solid regime or on surfaces~\cite{Gottwald-JCP-2016, Gottwald-JCPNote-2016}.
For these cases, it will be interesting to check if the obtained spectral density depends on whether the underlying explicit dynamics is performed classically or quantum-mechanically.
According to the results of this paper, the semiclassical hybrid schemes have been proven as too inaccurate for this purpose and another approximate method should be considered.
Nonetheless, if the spectral density turned out to be indeed independent on dynamical quantum effects also for realistic environments, this would imply that a classically calculated spectral density can be used in a reduced density matrix propagation which allows for a quantum treatment without any further approximations.
The spectral density would, thus, provide a useful link between robust \MD\ simulations and fully exact reduced quantum dynamics on the basis of an effective \CL\ model.

\bibliography{./semiclassics,./GLE,./own}

\end{document}